\documentclass[twocolumn,prl,showpacs,floatfix]{revtex4}

\usepackage{graphicx}
\usepackage{dcolumn}
\usepackage{bm}
\usepackage{amsmath}

\begin{document}

\title{
Bosonic molecules in rotating traps
}

\author{Igor Romanovsky}
\author{Constantine Yannouleas}
\author{Leslie O. Baksmaty}
\author{Uzi Landman}

\affiliation{School of Physics, Georgia Institute of Technology,
             Atlanta, Georgia 30332-0430}

\date{To appear in Phys. Rev. Lett.}

\begin{abstract}
We present a variational many-body wave function for repelling bosons in 
rotating traps, focusing on rotational frequencies that do not lead to
restriction to the lowest Landau level. This wave function incorporates 
correlations beyond the Gross-Pitaevskii (GP) mean field approximation, and it 
describes rotating boson molecules (RBMs) made of localized bosons that form 
polygonal-ring-like crystalline patterns in their intrinsic frame of reference. 
The RBMs exhibit characteristic periodic dependencies of the ground-state 
angular momenta on the number of bosons in the polygonal rings. For small 
numbers of neutral bosons, the RBM ground-state energies are found to be always 
lower than those of the corresponding GP solutions, in particular in the regime 
of GP vortex formation.  
\end{abstract}


\maketitle

Recent experimental advances in the field of trapped ultracold neutral
bosonic gases have enabled control of the strength of interatomic 
interactions over wide ranges \cite{cor,grei,par,wei}, from the very weak to the
very strong. This control is essential for experimental
realizations of novel states of matter beyond the well known Bose-Einstein
condensate \cite{grei,par,wei}. In this context, the {\it linear\/} 1D 
Tonks-Girardeau regime of impenetrable trapped bosons has generated intensive 
theoretical activity \cite{gir2,dunj} and several experimental realizations of 
it have been reported most recently \cite{par,wei}.

Here we address the properties of strongly-repelling impenetrable 
bosons in {\it rotating\/} ring-shaped or 2D harmonic traps. To this end, 
we recall that impenetrable bosons are ``localized'' relative to each 
other \cite{rom,wei} and exhibit nontrivial intrinsic crystalline 
correlations \cite{rom}. For a small number of bosons, $N$, these crystalline
arrangements are reminiscent of the structures exibited by the well-studied
rotating electron molecules (REMs) in quantum dots under high magnetic fields 
\cite{yl3,yl4}. Consequently, we use in the following the term rotating boson 
molecules (RBMs). A central result of our study is that the point-group
symmetries of the intrinsic crystalline structures give rise to characteristic 
regular patterns (see below) in the ground-state spectra and associated angular
momenta of the RBMs as a function of the rotational frequency for neutral
bosons (or the magnetic field for charged bosons).

An unexpected result of our studies is that the rotation of repelling bosons 
(even those interacting weakly) does not necessarily lead to formation of 
vortices, as is familiar from the case of rotating Bose-Einstein condenstates 
(BECs). In particular, for small $N$, we 
will show that the Gross-Pitaevskii energies (including those corresponding to
formation of vortices) remain always higher compared to the 
ground-state energies of the RBMs. Of course, we expect that the rotating BEC 
will become the preferred ground state for sufficiently large $N$
in the case of weakly repelling neutral bosons. We anticipate, however, that
it will be feasible to test our unexpected results for small $N$ by using
rotating optical lattices, where it is established that a small finite
number of atoms can be trapped per given site \cite{grei}. 

In a non-rotating trap, it is natural to describe a localized boson (at a 
position ${\bf R}_j$) by a simple displaced gaussian \cite{rom}. When the 
rotation of the trap is considered, the gaussian needs to be modified by a
phase factor, determined through the analogy between the one-boson Hamiltonian 
in the rotating frame of reference and the planal motion of a charged particle 
under the influence of a perpendicular magnetic field $B$ (described in the 
symmetric gauge). That is, the single-particle wave function of a localized 
boson is
\begin{equation} 
\varphi_j({\bf r})\equiv \varphi({\bf r},{\bf R}_j)=
\frac{1}{ \sqrt{\pi} \lambda } 
\exp \left[ 
\frac{ ({\bf r}-{\bf R}_j)^2 } {2\lambda^2} 
- i {\bf r} {\bf \cdot} ({\bf Q} \times {\bf R}_j) \right],
\label{disg}
\end{equation}
with  ${\bf Q} \equiv  {\bf \hat{z}} /(2\Lambda^2)$ 
and the width of the Gaussian $\lambda$ is a variational parameter;
$\Lambda \equiv l_B = \sqrt{\hbar c/(eB)}$
for the case of a perpendicular magnetic field ${\bf B}$, and 
$\Lambda \equiv l_\Omega = \sqrt{\hbar /(2m\Omega)}$
in the case of a rotating trap with rotational frequency ${\bf \Omega}$.
Note that we consider a 2D trap, so that ${\bf r} \equiv (x,y)$ and
${\bf R} \equiv (X,Y)$.
The hamiltonian corresponding to the single-particle kinetic energy is given by 
$H_K({\bf r})= ( {\bf P} - \hbar {\bf Q} \times {\bf r})^2/(2m)$,
for the case of a magnetic field, and by $H_K({\bf r})=
( {\bf P} - \hbar {\bf Q} \times {\bf r})^2/(2m) 
- m \Omega^2 {\bf r}^2/2$, for the case of a rotating frame of
reference \cite{note1}. 

A toroidal trap with radius $r_0$ can be specified by the confining 
potential
\begin{equation}
V({\bf r})= \frac{\hbar \omega_0}{2} (r-r_0)^n/l_0^n,
\label{pot}
\end{equation}
with $l_0=\sqrt{\hbar/(m\omega_0)}$ being the characteristic
length of the 2D trap. For $n \gg 2$ and 
$\l_0/r_0 \to 0$ this potential approaches the limit of a toroidal 
trap with zero width, which has been considered often in previous theoretical 
studies (see, e.g., Ref.\ \cite{kart}). 
In the following, we consider the case with $n=2$, which is
more realistic from the experimental point of view. In this case, in the  limit 
$r_0=0$, one recovers a harmonic trapping potential.
%
\begin{figure}[t]
\centering{\includegraphics[width=5.50cm]{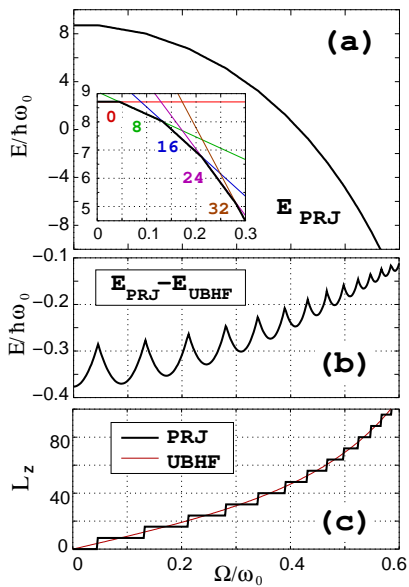}}
\caption{
Properties of $N=8$ neutral repelling bosons in a rotating toroidal 
trap as a function of the reduced rotational frequency $\Omega/\omega_0$.  
The confining potential is given by Eq.\ (\ref{pot}) with $n=2$ and radius 
$r_0=3 l_0$, and the interaction-strength parameter was chosen as $R_\delta=50$. 
(a) RBM ground-state energies, $E^{\text{PRJ}}$.
The inset shows the range $0 \leq \Omega/\omega_0 \leq 0.3$. The numbers denote
ground-state magic angular momenta.
(b) Energy difference $E^{\text{PRJ}}-E^{\text{UBHF}}$.
(c) Total angular momenta associated  with (i) the RBM ground states [thick solid
line (showing steps and marked as PRJ); online black] and (ii) 
the UBHF solutions (thin solid line; online red)
}
\end{figure}

To construct an RBM variational many-body wave function describing $N$ impenetrable
bosons in the toroidal trap, we use $N$ displaced orbitals 
$\varphi({\bf r}, {\bf R}_i)$, $i=1,2,...,N$ [see Eq.\ (\ref{disg})] centered at 
the vertices of a regular polygon. Then, we first construct an 
{\it unrestricted\/} Bose Hartree-Fock (UBHF) permanent \cite{rom}
$|\Phi^{\text{UBHF}}_N \rangle 
\propto \sum_{P(i_m)} \varphi_1({\bf r}_{i_1}) \varphi_2({\bf r}_{i_2}) ...
\varphi_N({\bf r}_{i_N})$. The UBHF permanent breaks the circular symmetry of 
the many-body hamiltonian. The ``symmetry dilemma'' is resolved through a 
subsequent ``symmetry-restoration'' step accomplished via projection techniques 
\cite{yl4,lpy,yl1},
i.e., we construct a many-body wave function with good total angular
momentum by applying the projection operator   
$\hat{\cal P}_L = (1/2\pi) \int_0^{2\pi} d\theta \exp[i \theta (L-\hat{L})]$,
so that the final RBM wave function is given by
\begin{equation}
|\Psi^{\text{PRJ}}_{N,L} \rangle =
\frac{1}{2 \pi} \int_0^{2\pi} 
d\theta |\Phi^{\text{UBHF}}_N (\theta) \rangle e^{i\theta L}.
\label{wfprj}
\end{equation}
$|\Phi^{\text{UBHF}}_N (\theta) \rangle$ is the original UBHF permanent rotated 
by an azimuthal angle $\theta$. We note that, in addition to having
good angular momenta, the projected (PRJ) wave function 
$|\Psi^{\text{PRJ}}_{N,L} \rangle$ has also a {\it lower\/} energy than that 
of $|\Phi^{\text{UBHF}}_N \rangle$
[see, e.g. $E_L^{\text{PRJ}} - E^{\text{UBHF}}$ in Fig.\ 1(b)]. 
The projected ground-state energy is given by
\begin{equation}
E_L^{\text{PRJ}} = \int_0^{2\pi} h(\theta) e^{i\theta L} d\theta \left / 
\int_0^{2\pi} n(\theta) e^{i\theta L} \right . d\theta,
\label{elprj}
\end{equation} 
where $h(\theta)= \langle \Phi_N^{\text{UBHF}}(\theta=0)  | H |
\Phi_N^{\text{UBHF}}(\theta)\rangle$ and
$n(\theta)= \langle \Phi_N^{\text{UBHF}}(\theta=0)  |
\Phi_N^{\text{UBHF}}(\theta)\rangle$; the latter term ensures proper
normalization. 

The many-body hamiltonian is given by 
${\cal H}=\sum_{i=1}^N [H_K({\bf r}_i)+V({\bf r}_i)] + \sum_{i<j}^N 
V({\bf r}_i,{\bf r}_j)$, with the interparticle interaction being given
by a contact potential $V_\delta=g \delta ({\bf r}_i - {\bf r}_j)$
for neutral bosons and a Coulomb potential $V_C=Z^2e^2/|{\bf r}_i - {\bf r}_j|$
for charged bosons. The parameter that controls the strength of the
interparticle repulsion relative to the zero-point kinetic energy is given
by $R_\delta=gm/(2\pi\hbar^2)$ \cite{rom} for a contact potential and 
$R_W=Z^2e^2/(\hbar \omega_0 l_0)$ \cite{rom,yl10} for a Coulomb
repulsion.

For a given value of the dimensionless rotational frequency,
$\Omega/\omega_0$, the projection yields wave functions and energies
for a whole rotational band comprising many angular momenta. In the following, 
we focus on the ground-state wave function (and corresponding angular momentum 
and energy) associated with the lowest energy in the band.  

Fig.\ 1(a) displays the ground-state energy $E_{\text{PRJ}}$ of $N=8$ bosons
in a toroidal trap as a function of
the dimensionless rotational frequency $\Omega/\omega_0$, with $\omega_0$
being the trap frequency. The prominent features in Fig.\ 1(a) are: (i) the 
energy diminishes as $\Omega/\omega_0$ increases; this is an effect of the
centrifugal force, and (ii) the $E_{\text{PRJ}}$ curve consists of linear 
segments, each one associated with a given angular momentum $L$. Most remarkable
is the regular variation of the values of $L$ with a constant step of $N$
units (here $N=8$) [see inset in Fig.\ 1(a) and Fig.\ 1(c)]. These preferred
angular momenta $L=kN$ with integer $k$, are reminiscent of the 
so called ``magic angular momenta'' familiar from studies of electrons under
high-magnetic fields in 2D semiconductor quantum dots \cite{yl3,yl4}.
%
\begin{figure}[t]
\centering{\includegraphics[width=5.5cm]{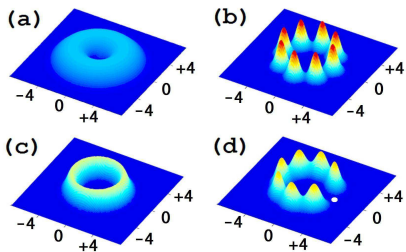}}
\caption{
Single-particle densities and CPDs for $N=8$ bosons in a rotating toroidal trap
with $\Omega/\omega_0=0.2$ and $R_\delta=50$. The remaining trap parameters
are as in Fig.\ 1. (a) GP sp-density. (b) UBHF sp-density exhibiting
breaking of the circular symmetry. (c) RBM sp-density exhibiting circular
symmetry. (d) CPD for the RBM wave function [PRJ wave function, see 
Eq.\ (\ref{wfprj})] revealing the hidden point-group symmetry in the
intrinsic frame of reference. The observation point is denoted by a white dot.
The RBM ground-state angular momentum is $L_z=16$. Lengths in units of
$l_0$. The vertical scale is the same for (b), (c), and (d), but
different for (a).
}
\end{figure}
%
The preferred angular momenta reflect the intrinsic molecular structure
of the localized impenetrable bosons. We note, that the (0,8) polygonal-ring 
arrangement is obvious in the single-particle density associated with
the UBHF permanent [see Fig. 2(b)]; (0,8) denotes no particles in the inner
ring and 8 particles in the outer one. After restoration of symmetry, however,
the single-particle (sp) density is circularly symmetric [see the PRJ sp-density
in Fig.\ 2(c)] and the intrinsic crystallinity becomes ``hidden''; 
it can, however, be revealed via the conditional probability distribution 
\cite{rom,yl4} [CPD, see Fig.\ 2(d)]. We note the Gross-Pitaevskii
sp-density in Fig.\ 2(a), which is clearly different from the PRJ
density in Fig.\ 2(c). 

The internal structure for {\it charged\/} bosons in a toroidal trap (not
shown) is similar to that of neutral bosons (Fig.\ 2), i.e., a (0,8) ring 
arrangement, portrayed also in the stepwise variation (in steps of 8 units)
of the total angular momenta. The internal structure is also
reflected in the variation of the ground-state total energy as a function of
the magnetic field. In contrast to the case of neutral bosons,
however, the ground-state energy curve for charged bosons is not composed of 
linear segments, but of intersecting inverted-parabola-type pieces; this is due 
to the positive contribution of the Lorentz force compared to the negative 
contribution of the centrifugal force in a rotating trap.
%
\begin{figure}[b]
\centering{\includegraphics[width=5.50cm]{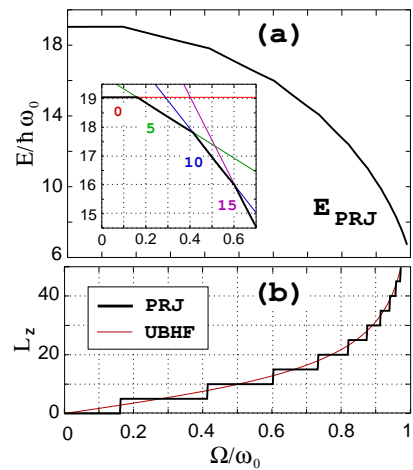}}
\caption{
Properties of $N=6$ neutral bosons in a rotating harmonic 
trap as a function of the reduced rotational frequency $\Omega/\omega_0$.
The confining potential is given by Eq.\ (\ref{pot}) with $n=2$ and 
$r_0=0$, and the interaction-strength parameter was chosen as $R_\delta=50$.
The intrinsic molecular structure is $(1,5)$.
(a) RBM ground-state energies, $E^{\text{PRJ}}$. The inset shows a smaller
range. The numbers denote ground-state angular momenta.
(b) Total angular momenta associated  with (i) the RBM ground states (thick solid
line showing steps; online black) and (ii) the UBHF solutions (thin solid line; 
online red).
}
\end{figure}

For RBMs in rotating {\it harmonic\/} traps,
the polygonal-ring pattern of localized bosons becomes more
complex than the simple $(0,N)$ arrangement that appears naturally in
a toroidal trap. Indeed, in harmonic traps, one anticipates the emergence of
concentric ring structures. For $N=6$ neutral bosons in a harmonic trap,
we observe that, as in the case of a toroidal trap, the ground-state energy as a
function of the reduced rotational frequency, $\Omega/\omega_0$, [Fig.\ 3(a)] 
is composed of linear segments, but now the corresponding magic angular
momenta [Fig.\ 3(b)] vary in steps of $N-1=5$ units. This indicates an
RBM consisting of {\it two\/} polygonal rings; denoted as a $(1,5)$ structure, 
with the inner ring having a single boson and the outer ring five.
%
\begin{figure}[t]
\centering{\includegraphics[width=5.50cm]{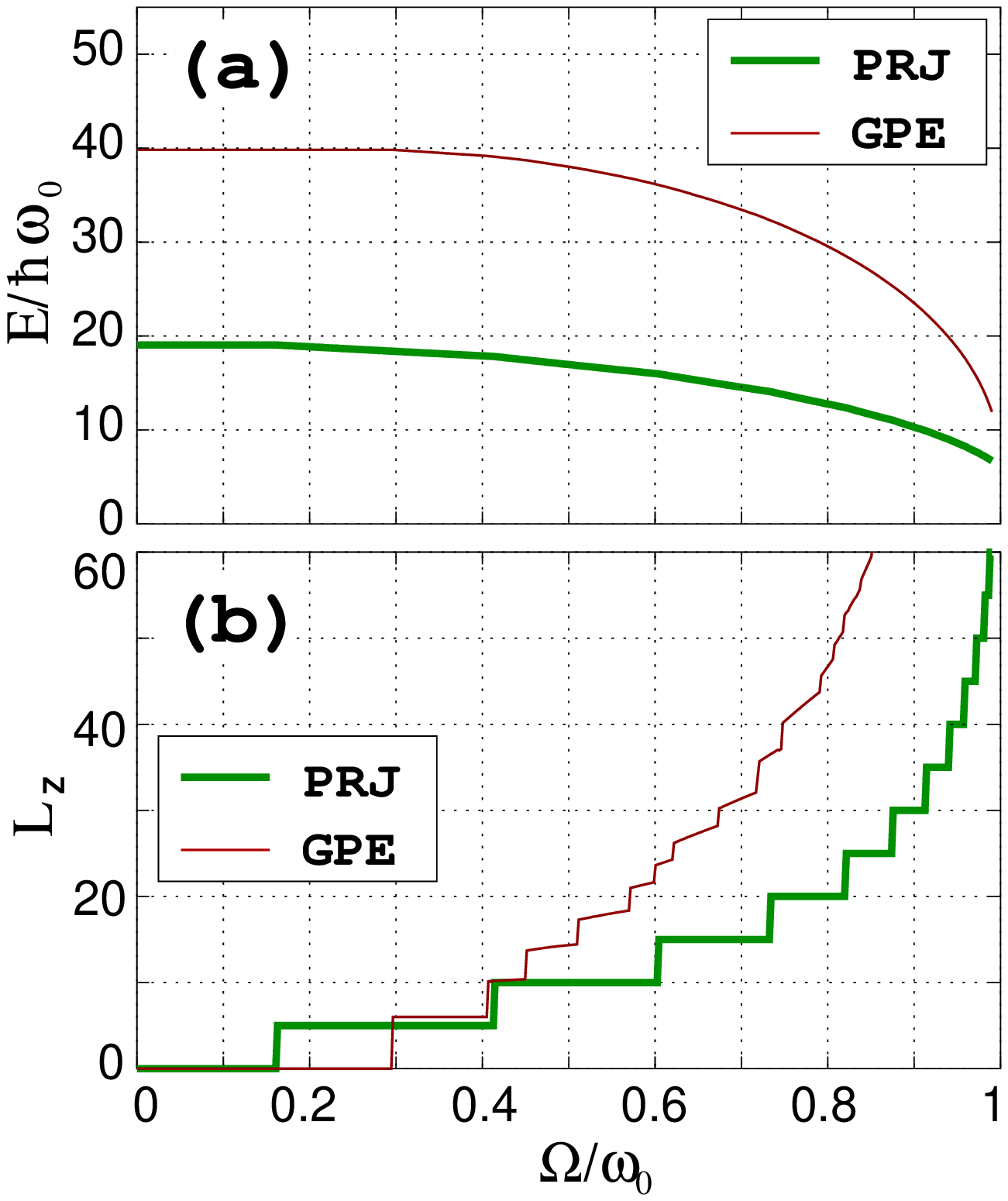}}
\centering{\includegraphics[width=7.00cm]{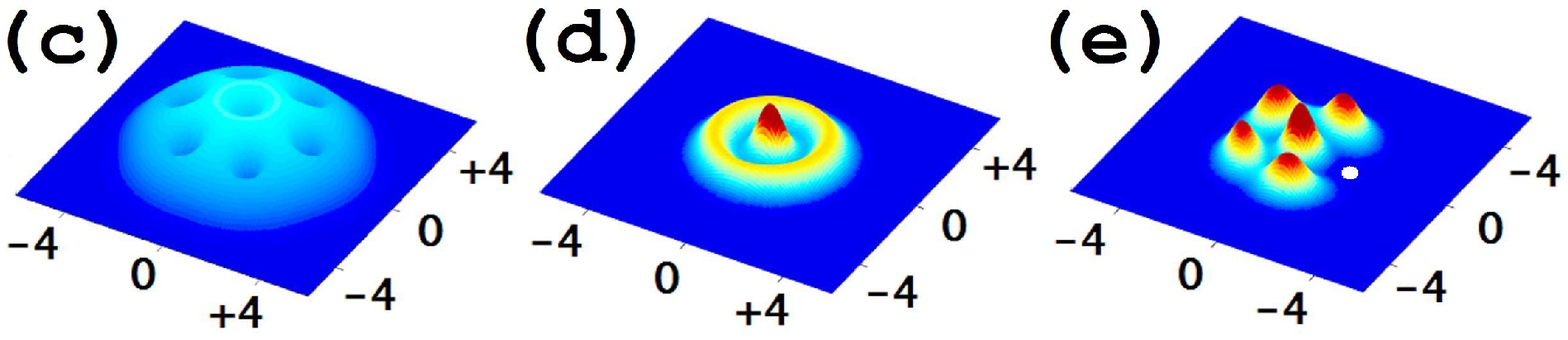}}
\caption{
Properties of GP solutions (thin solid line; online red) versus those of RBM 
wave functions (thick solid line; online green) for $N=6$ neutral bosons 
as a function of the reduced rotational frequency $\Omega/\omega_0$. A harmonic 
trap is considered, and the interaction strength equals $R_\delta=50$.
(a) Ground-state energies. (b) Associated ground-state angular momenta.
(c) GP (BEC) sp-density at $\Omega/\omega_0=0.65$ having 7 vortices
with a 6-fold symmetry (thus exhibiting breaking of the circular symmetry).
(d) RBM sp-density at $\Omega/\omega_0=0.65$ which does not break the
circular symmetry. (e) CPD of the RBM at $\Omega/\omega_0=0.65$ revealing the
intrinsic (1,5) crystalline pattern. The white dot denotes the observation
point ${\bf r}_0$. Note the dramatic difference in spatial extent between
the GP and RBM wave functions [compare (c) with (d) and (e). Lengths in units 
of $l_0$. The vertical scale is the same for (d) and (e), but different for (c).
}
\end{figure}

In Fig.\ 4(a), we display the RBM and mean-field GP ground-state
energies of $N=6$ strongly repelling (i.e., $R_\delta=50$) neutral bosons in a 
harmonic trap as a function of the reduced angular frequency of the trap.
The GP curve (thin solid line; online red) remains well above the RBM curve 
(thick solid line; online green) in the whole range 
$ 0 \leq \Omega/\omega_0 \leq 1$. The RBM ground-state angular momenta exhibit 
again the periodicity in steps of five units [Fig.\ 4(b)].
As expected, the GP total angular momenta
are quantized [$L_z=0$ (no-vortex) or $L_z=6$ (one central vortex)] only for
an initial range $0 \leq \Omega/\omega_0 \leq 0.42$. For 
$\Omega/\omega_0 \geq 0.42$, the GP total angular momentum takes non-integer
values and ceases to be a good quantum number, reflecting the broken-symmetry
character of the associated mean field, with each kink signaling the
appearance of a different vortex pattern of $p$-fold symmetry ($p=1,2,3,4,...$)
\cite{butt}; see an example in Fig.\ 4(c).
%
\begin{figure}[t]
\centering{\includegraphics[width=5.50cm]{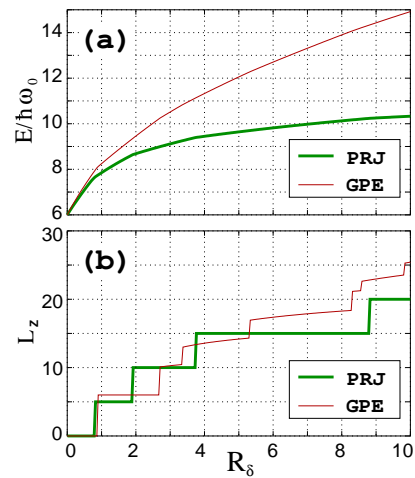}}
\caption{
Properties of GP solutions (thin solid line; online red) versus those of RBM wave
functions (thick solid line; online green) for $N=6$ bosons as a function of the 
interaction strength $R_\delta$. A harmonic trap is considered, and the reduced 
rotational frequency equals $\Omega/\omega_0=0.85$.
(a) Ground-state energies (b) Associated ground-state angular momenta.
}
\end{figure}

The energetic superiority of the RBM wave function over the GP solution
demonstrated in Fig.\ 4(a) was to be expected, since we considered the case
of strongly repelling bosons. Unexpectedly, however, for a small number of
neutral bosons the energetic advantage of the RBM persists even for weakly 
repelling bosons, as illustrated in Fig.\ 5(a). Indeed, Fig.\ 5(a) displays the 
RBM (thick solid line; online green) and  GP (thin solid line; online red) 
ground-state energies for $N=6$ neutral bosons in a trap rotating with 
$\Omega/\omega_0=0.85$ as a function of the interaction parameter $R_\delta$. 
The surprising result in Fig.\ 5(a) is that the GP energy remains above the RBM 
curve even for $R_\delta \rightarrow 0$. Of course the RBM wave function is very 
close to that of a BEC without vortices when $R_\delta \rightarrow 0$ 
(BECs {\it without\/}  vortices are approximately feasible for small $N$). 
However, for small $N$, our results show that BECs {\it with vortices\/} 
(i.e., for $L_z \geq N$) are not the preferred many-body ground states; instead, 
formation of RBMs is favored.  
Note that the energy difference $E^{\text{GP}}-E^{\text{PRJ}}$ increases rapidly
with increasing $R_\delta$, reflecting the fact that the RBM energies saturate 
(as is to be expected from general arguments), while the GP energies (even with
vortices fully accounted for) exhibit an unphysical divergence as 
$R_\delta \rightarrow \infty$ [Fig. 5(a)]; we have checked this trend up to
values of $R_\delta=100$ (not shown). 
Of interest again is the different behavior of the RBM and GP ground state
angular momenta [Fig.\ 5(b)] (see also discussion of Fig.\ 4(b)).

In conclusion, we have introduced (and studied the ground-state properties of) 
a variational many-body wave function for repelling bosons in rotating traps 
that incorporates correlations beyond the Gross-Pitaevskii mean-field 
approximations. This variational wave function describes rotating boson 
molecules, i.e., localized bosons arranged in polygonal-ring-type patterns in 
their intrinsic frame of reference. For small numbers of neutral bosons, and in 
particular in the case of GP vortex formation, the RBM ground-state energies are
lower than those associated with the corresponding Gross-Pitaevskii BEC
solutions. Given the large differences between the properties of the RBM and
BEC wave functions (which become more pronounced for larger interaction 
parameter $R_\delta$), and the recently demonstrated ability to experimentally
control $R_\delta$ \cite{cor,grei,par,wei}, we anticipate that our results could
be tested in experiments involving rotating optical lattices. 
Detection of RBMs could be based on a variety of approaches \cite{note2}, 
such as the measurement of the spatial extent [contrast the RBM and BEC spatial 
extents in Figs.\ 4(c)-4(e)], or the use of Hanbury Brown-Twiss-type 
experiments \cite{brtw} to directly detect the intrinsic crystalline structure 
of the RBM.

This work is supported by the US D.O.E. (Grant No. FG05-86ER45234) and the NSF
(Grant No. DMR-0205328).


\begin{thebibliography}{99}
\bibitem{cor}
S.L. Cornish {\it et al.\/},
Phys. Rev. Lett. {\bf 85}, 1795 (2000).
\bibitem{grei}
M. Greiner {\it et al.\/},
Nature (London) {\bf 415}, 39 (2002).
\bibitem{par}
B. Paredes {\it et al.\/},
Nature {\bf 429}, 277 (2004).
\bibitem{wei}
G.T. Kinoshita, T. Wenger, and D.S. Weiss,
Science {\bf 305}, 1125 (2004),
\bibitem{gir2}
M.D. Girardeau and E.M. Wright,
Laser Physics {\bf 12}, 8 (2002).
\bibitem{dunj}
V. Dunjko {\it et al.\/},
Phys. Rev. Lett. {\bf 86}, 5413 (2001).
\bibitem{rom}
I. Romanovsky, C. Yannouleas, and U. Landman,
Phys. Rev. Lett. {\bf 93}, 230405 (2004).
\bibitem{yl3}
C. Yannouleas and U. Landman,
Phys. Rev. B {\bf 68}, 035326 (2003).
\bibitem{yl4}
C. Yannouleas and U. Landman,
Phys. Rev. B {\bf 70}, 235319 (2004);
YS Li, C. Yannouleas, and U. Landman,
Phys. Rev. B {\bf 73}, 075301 (2006).
\bibitem{note1}
The single-particle wave function in Eq.\ (\ref{disg}) and 
the many-body projected wave function in Eq. (\ref{wfprj}) contain 
contributions from higher Landau levels. These wave functions belong
exclusively to the lowest Landau level (LLL) only in the
limit when $\lambda=\sqrt{2} l_B$ in the case of a magnetic field,
or $\lambda=\sqrt{2} l_\Omega$ and $\Omega/\omega_0=1$ in the
case of a rotating trap. LLL investigations of neutral bosons using exact 
diagonalization (EXD) techniques have attempted to introduce analogies with the
liquid-like bosonic Laughlin and composite-fermion wave functions [see e.g.,
Th. Jolicoeur and N. Regnault, Phys. Rev. B {\bf 70}, 241307 (2004)].
Recently, however, the possibility of localization of neutral bosons in the LLL 
is also attracting attention. In particular, the EXD study in Ref.\ \cite{note2}
shows that, for small $N$, formation of vortices in the LLL is not a prevalent
phenomenon, and their appearance ``is restricted to the vicinity of some
critical values of the rotational frequency ...'' 
\bibitem{kart}
P.F. Kartsev,
Phys. Rev. A {\bf 68}, 063613 (2003).
\bibitem{lpy}
P.-O. L\"{o}wdin,
Rev. Mod. Phys. {\bf 34}, 520 (1962);
R.E. Peierls and J. Yoccoz,
Proc. Phys. Soc. London, Sect. A {\bf 70}, 381 (1957).
\bibitem{yl1}
C. Yannouleas and U. Landman,
J. Phys.: Condens. Matter {\bf 14}, L591 (2002);
Phys. Rev. B {\bf 66}, 115315 (2002).
\bibitem{yl10}
C. Yannouleas and U. Landman,
Phys. Rev. Lett. {\bf 82}, 5325 (1999).
\bibitem{butt}
R.A. Butts and D.S. Rokhsar,
Nature (London) {\bf 397}, 327 (1999). 
\bibitem{note2}
A more detailed discussion on possible detection approaches of strongly
correlated bosonic states is given in N. Barberan {\it et al.\/},
Phys. Rev. A {\bf 73}, 063623 (2006).
\bibitem{brtw}
M. Schellekens {\it et al.\/},
Science {\bf 310}, 648 (2005).
\end{thebibliography}
\end{document}